\documentclass{sig-alternate}
\usepackage{xspace,latexsym,amssymb,amsmath,verbatim}

\newtheorem{definition}{Definition}
\newtheorem{fact}{Fact}

\newcommand{\tuple}[1]{\ensuremath{\left\langle #1 \right\rangle}}
\newcommand{\mypara}[1]{\vspace*{0.1in}\noindent\textbf{#1} \xspace}

\newcommand{\SB}{\ensuremath{\mathit {SB}}\xspace}

\title{Slicing: A New Approach to Privacy Preserving\\Data Publishing}

\begin{document}

\numberofauthors{1}

\author{
%
%
\alignauthor
Tiancheng Li, Ninghui Li, Jian Zhang, Ian Molloy\vspace*{.05in}\\
       \affaddr{Purdue University, West Lafayette, IN 47907}\\
       \email{\{li83,ninghui\}@cs.purdue.edu, jianzhan@purdue.edu, imolloy@cs.purdue.edu}
}

\maketitle \thispagestyle{empty}
\begin{abstract}
Several anonymization techniques, such as generalization and bucketization, have been designed for
privacy preserving microdata publishing. Recent work has shown that generalization loses
considerable amount of information, especially for high-dimensional data. Bucketization, on the
other hand, does not prevent membership disclosure and does not apply for data that do not have a
clear separation between quasi-identifying attributes and sensitive attributes.

In this paper, we present a novel technique called slicing, which partitions the data both
horizontally and vertically. We show that slicing preserves better data utility than generalization
and can be used for membership disclosure protection. Another important advantage of slicing is
that it can handle high-dimensional data.
We show how slicing can be used for attribute disclosure protection and develop an efficient
algorithm for computing the sliced data that obey the $\ell$-diversity requirement.
Our workload experiments confirm that slicing preserves better utility than generalization and is
more effective than bucketization in workloads involving the sensitive attribute. Our experiments
also demonstrate that slicing can be used to prevent membership disclosure.
\end{abstract}
\sloppypar


\section{Introduction}
\label{sec:introduction}

%
%

Privacy-preserving publishing of microdata has been studied extensively in recent years.
\emph{Microdata} contains records each of which contains information about an individual
entity, such as a person, a household, or an organization.
%
%
%
%
%
%
Several microdata anonymization techniques have been proposed.  The most popular ones are
generalization~\cite{Sam01,Swe02a} for $k$-anonymity~\cite{Swe02a} and
bucketization~\cite{XT06b,MKM+07,KSY+07} for $\ell$-diversity~\cite{MGK+06}. In both approaches,
attributes are partitioned into three categories: (1) some attributes are {\em identifiers} that
can uniquely identify an individual, such as {\em Name} or {\em Social Security Number}; (2) some
attributes are \emph{Quasi-Identifiers (QI)}, which the adversary may already know (possibly from
other publicly-available databases) and which, when taken together, can potentially identify an
individual, e.g., \emph{Birthdate}, \emph{Sex}, and \emph{Zipcode}; (3) some attributes are
\emph{Sensitive Attributes (SAs)}, which are unknown to the adversary and are considered sensitive,
such as {\em Disease} and {\em Salary}.

In both generalization and bucketization, one first removes identifiers from the data and then
partitions tuples into buckets. The two techniques differ in the next step. Generalization
transforms the QI-values in each bucket into ``less specific but semantically consistent'' values
so that tuples in the same bucket cannot be distinguished by their QI values. In bucketization, one
separates the SAs from the QIs by randomly permuting the SA values in each bucket. The anonymized
data consists of a set of buckets with permuted sensitive attribute values.

\subsection{Motivation of Slicing}
It has been shown~\cite{Agg05,KG06,XT06b} that generalization for $k$-anonymity losses
considerable amount of information, especially for high-dimensional data. This is due to the
following three reasons.
First, generalization for $k$-anonymity suffers from the curse of dimensionality. In order for
generalization to be effective, records in the same bucket must be close to each other so that
generalizing the records would not lose too much information. However, in high-dimensional data,
most data points have similar distances with each other, forcing a great amount of generalization
to satisfy $k$-anonymity even for relative small $k$'s.
Second, in order to perform data analysis or data mining tasks on the generalized table,
the data analyst has to make the uniform distribution assumption that every value in a
generalized interval/set is equally possible, as no other distribution assumption can be
justified. This significantly reduces the data utility of the generalized data.
%
%
Third, because each attribute is generalized separately, correlations between different attributes
are lost. In order to study attribute correlations on the generalized table, the data analyst has
to assume that every possible combination of attribute values is equally possible. This is an
inherent problem of generalization that prevents effective analysis of attribute correlations.

While bucketization~\cite{XT06b,MKM+07,KSY+07} has better data utility than generalization, it has
several limitations. First, bucketization does not prevent membership disclosure~\cite{NAC07}.
Because bucketization publishes the QI values in their original forms, an adversary can find out
whether an individual has a record in the published data or not.
As shown in~\cite{Swe02a}, 87\% of the individuals in the United States can be uniquely identified
using only three attributes (\emph{Birthdate}, \emph{Sex}, and \emph{Zipcode}). A microdata (e.g.,
census data) usually contains many other attributes besides those three attributes. This means that
the membership information of most individuals can be inferred from the bucketized table.
Second, bucketization requires a clear separation between QIs and SAs. However, in many datasets,
it is unclear which attributes are QIs and which are SAs.
Third, by separating the sensitive attribute from the QI attributes, bucketization breaks the
attribute correlations between the QIs and the SAs.
%
%

In this paper, we introduce a novel data anonymization technique called {\em slicing} to improve
the current state of the art. Slicing partitions the dataset both vertically and horizontally.
Vertical partitioning is done by grouping attributes into columns based on the correlations among
the attributes. Each column contains a subset of attributes that are highly correlated. Horizontal
partitioning is done by grouping tuples into buckets. Finally, within each bucket, values in each
column are randomly permutated (or sorted) to break the linking between different columns.

The basic idea of slicing is to break the association cross columns, but to preserve the
association within each column.  This reduces the dimensionality of the data and preserves better
utility than generalization and bucketization.  Slicing preserves utility because it groups
highly-correlated attributes together, and preserves the correlations between such attributes.
Slicing protects privacy because it breaks the associations between uncorrelated attributes, which
are infrequent and thus identifying. Note that when the dataset contains QIs and one SA,
bucketization has to break their correlation; slicing, on the other hand, can group some QI
attributes with the SA, preserving attribute correlations with the sensitive attribute.

The key intuition that slicing provides privacy protection is that the slicing process ensures that
for any tuple, there are generally multiple matching buckets.
Given a tuple $t=\tuple{v_1,v_2,\ldots,v_c}$, where $c$ is the number of columns, a bucket is a
matching bucket for $t$ if and only if for each $i$ ($1\leq i\leq c$), $v_i$ appears at least once
in the $i$'th column of the bucket. Any bucket that contains the original tuple is a matching
bucket.  At the same time, a matching bucket can be due to containing other tuples each of which
contains some but not all $v_i$'s.

\subsection{Contributions \& Organization} In this paper, we present a novel technique called {\em slicing} for privacy-preserving data publishing. Our contributions include the following.

First, we introduce slicing as a new technique for privacy preserving data publishing.  Slicing has
several advantages when compared with generalization and bucketization. It preserves better data
utility than generalization. It preserves more attribute correlations with the SAs than
bucketization. It can also handle high-dimensional data and data without a clear separation of QIs
and SAs.


Second, we show that slicing can be effectively used for preventing attribute disclosure, based on
the privacy requirement of $\ell$-diversity. We introduce a notion called $\ell$-diverse slicing,
which ensures that the adversary cannot learn the sensitive value of {\em any} individual with a
probability greater than $1/\ell$.

Third, we develop an efficient algorithm for computing the sliced table that satisfies
$\ell$-diversity. Our algorithm partitions attributes into columns, applies column generalization,
and partitions tuples into buckets. Attributes that are highly-correlated are in the same column;
this preserves the correlations between such attributes. The associations between uncorrelated
attributes are broken; the provides better privacy as the associations between such attributes are
less-frequent and potentially identifying.

Fourth, we describe the intuition behind membership disclosure and explain how slicing prevents
membership disclosure. A bucket of size $k$ can potentially match $k^c$ tuples where $c$ is the
number of columns. Because only $k$ of the $k^c$ tuples are actually in the original data, the
existence of the other $k^c-k$ tuples hides the membership information of tuples in the original
data.

Finally, we conduct extensive workload experiments. Our results confirm that slicing preserves much better data utility than generalization. In workloads involving the sensitive attribute, slicing is also more effective than bucketization. In some classification experiments, slicing shows better
performance than using the original data (which may overfit the model). Our experiments also show the limitations of bucketization in membership disclosure protection and slicing remedies these limitations.

The rest of this paper is organized as follows. In Section~\ref{sec:slicing}, we formalize the
slicing technique and compare it with generalization and bucketization. 
We define $\ell$-diverse slicing for attribute disclosure protection in Section~\ref{sec:attribute} and develop an efficient algorithm to achieve $\ell$-diverse slicing in Section~\ref{sec:algorithms}. In Section~\ref{sec:membership}, we explain how slicing prevents membership disclosure. Experimental results are presented in Section~\ref{sec:experiments} and related work is discussed in Section~\ref{sec:related}. We conclude the paper and discuss future research in Section~\ref{sec:conclusions}.


\section{Slicing}
\label{sec:slicing}
\begin{table*}
\centering
\begin{tabular}{ccc}
\centering
\begin{minipage}{2.1in}
\begin{tabular}{|c|c|c|c|}
\hline
Age&Sex&Zipcode&Disease\\
\hline
22&M&47906& dyspepsia\\
\hline
22&F&47906& flu\\
\hline
33&F&47905& flu\\
\hline
52&F&47905& bronchitis\\
\hline
54&M&47302& flu\\
\hline
60&M&47302& dyspepsia\\
\hline
60&M&47304& dyspepsia\\
\hline
64&F&47304& gastritis\\
\hline
\end{tabular}
\end{minipage}
&
\begin{minipage}{2.3in}
\begin{tabular}{|c|c|c|c|}
\hline
Age&Sex&Zipcode&Disease\\
\hline
$[$20-52$]$&*&4790*& dyspepsia\\
$[$20-52$]$&*&4790*& flu\\
$[$20-52$]$&*&4790*& flu\\
$[$20-52$]$&*&4790*& bronchitis\\
\hline
$[$54-64$]$&*&4730*& flu\\
$[$54-64$]$&*&4730*& dyspepsia\\
$[$54-64$]$&*&4730*& dyspepsia\\
$[$54-64$]$&*&4730*& gastritis\\
\hline
\end{tabular}
\end{minipage}
\begin{minipage}{2.5in}
\begin{tabular}{|c|c|c||c|}
\hline
Age&Sex&Zipcode&Disease\\
\hline
22&M&47906& flu\\
22&F&47906& dyspepsia\\
33&F&47905& bronchitis\\
52&F&47905& flu\\
\hline
54&M&47302& gastritis\\
60&M&47302& flu\\
60&M&47304& dyspepsia\\
64&F&47304& dyspepsia\\
\hline
\end{tabular}
\vspace*{.05in}
\end{minipage}
\\
(a) The original table&
\hspace*{-2.3in}(b) The generalized table&
\hspace*{-2.3in}(c) The bucketized table
\\
\end{tabular}

\vspace*{0.05in}
\begin{tabular}{ccc}
\centering
\begin{minipage}{3.0in}
\begin{tabular}{|c|c|c|c|}
\hline
Age&Sex&Zipcode&Disease\\
\hline
22:2,33:1,52:1&M:1,F:3&47905:2,47906:2& dysp.\\
22:2,33:1,52:1&M:1,F:3&47905:2,47906:2& flu\\
22:2,33:1,52:1&M:1,F:3&47905:2,47906:2& flu\\
22:2,33:1,52:1&M:1,F:3&47905:2,47906:2& bron.\\
\hline
54:1,60:2,64:1&M:3,F:1&47302:2,47304:2& flu\\
54:1,60:2,64:1&M:3,F:1&47302:2,47304:2& dysp.\\
54:1,60:2,64:1&M:3,F:1&47302:2,47304:2& dysp.\\
54:1,60:2,64:1&M:3,F:1&47302:2,47304:2& gast.\\
\hline
\end{tabular}
\end{minipage}
&
\begin{minipage}{2.0in}
\begin{tabular}{||c||c||c||c||}
\hline
Age&Sex&Zipcode&Disease\\
\hline
22&F&47906& flu\\
22&M&47905& flu\\
33&F&47906& dysp.\\
52&F&47905& bron.\\
\hline
54&M&47302& dysp.\\
60&F&47304& gast.\\
60&M&47302& dysp.\\
64&M&47304& flu\\
\hline
\end{tabular}
\end{minipage}
&
\begin{minipage}{2.2in}
\begin{tabular}{||c||c||}
\hline
(Age,Sex)&(Zipcode,Disease)\\
\hline
(22,M)&(47905,flu)\\
(22,F)&(47906,dysp.)\\
(33,F)&(47905,bron.)\\
(52,F)&(47906,flu)\\
\hline
(54,M)&(47304,gast.)\\
(60,M)&(47302,flu)\\
(60,M)&(47302,dysp.)\\
(64,F)&(47304,dysp.)\\
\hline
\end{tabular}
\vspace*{.05in}
\end{minipage}
\\
(d) Multiset-based generalization & (e) One-attribute-per-column slicing & (f) The
sliced table
\end{tabular}
\caption{An original microdata table and its anonymized versions using various anonymization
techniques}
\vspace*{-.2in}
\label{table:example}
\end{table*}


In this section, we first give an example to illustrate slicing. We then formalize slicing, compare
it with generalization and bucketization, and discuss privacy threats that slicing can address.

Table~\ref{table:example} shows an example microdata table and its anonymized versions using
various anonymization techniques.  The original table is shown in Table~\ref{table:example}(a). The
three QI attributes are $\{\mathit{Age}, \mathit{Sex}, \mathit{Zipcode}\}$, and the sensitive
attribute SA is $\mathit{Disease}$. A generalized table that satisfies $4$-anonymity is shown in
Table~\ref{table:example}(b), a bucketized table that satisfies $2$-diversity is shown in
Table~\ref{table:example}(c), a generalized table where each attribute value is replaced with the
the multiset of values in the bucket is shown in Table~\ref{table:example}(d), and two sliced
tables are shown in Table~\ref{table:example}(e) and ~\ref{table:example}(f).
%
%

Slicing first partitions attributes into columns. Each column contains a subset of attributes.
This vertically partitions the table. For example, the sliced table in
Table~\ref{table:example}(f) contains $2$ columns: the first column contains $\{\mathit{Age},
\mathit{Sex}\}$ and the second column contains $\{\mathit{Zipcode}, \mathit{Disease}\}$. The
sliced table shown in Table~\ref{table:example}(e) contains $4$ columns, where each column
contains exactly one attribute.

Slicing also partition tuples into buckets. Each bucket contains a subset of tuples. This
horizontally partitions the table. For example, both sliced tables in Table~\ref{table:example}(e)
and Table~\ref{table:example}(f) contain $2$ buckets, each containing $4$ tuples.

Within each bucket, values in each column are randomly permutated to break the linking between
different columns. For example, in the first bucket of the sliced table shown in
Table~\ref{table:example}(f), the values $\{(22,M), (22,F), (33,F), (52, F)\}$ are randomly
permutated and the values $\{(47906,\mathit{dyspepsia}),$ $(47906,\mathit{flu}),$
$(47905,\mathit{flu}),$ $(47905, \mathit{bronchitis})\}$ are randomly permutated so that the
linking between the two columns within one bucket is hidden.


\subsection{Formalization of Slicing}
\label{subsection:formalization}

Let $T$ be the microdata table to be published. $T$ contains $d$ attributes: $A=\{A_1,
A_2,\ldots,A_d\}$ and their attribute domains are $\{D[A_1],D[A_2],\ldots,D[A_d]\}$. A tuple $t\in
T$ can be represented as $t=(t[A_1],t[A_2],...,t[A_d])$ where $t[A_i]$ $(1\leq i\leq d)$ is the
$A_i$ value of $t$.
%

\begin{definition}[Attribute partition and columns]
An {\bf attribute partition} consists of several subsets of $A$, such that each attribute belongs
to exactly one subset. Each subset of attributes is called a {\bf column}. Specifically, let there
be $c$ columns $C_1,C_2,\ldots,C_c$, then $\cup_{i=1}^cC_i=A$ and for any $1\leq i_1\ne i_2\leq c$,
$C_{i_1}\cap C_{i_2}=\emptyset$.
\end{definition}

For simplicity of discussion, we consider only one sensitive attribute $S$. If the data contains multiple sensitive attributes, one can either consider them separately or consider their joint distribution~\cite{MGK+06}.
Exactly one of the $c$ columns contains $S$. Without loss of generality, let the column that contains $S$ be the last column $C_c$. This column is also called the {\em sensitive column}. All other columns $\{C_1,C_2,\ldots,C_{c-1}\}$ contain only QI attributes.

\begin{definition}[Tuple partition and buckets]
A {\bf tuple partition} consists of several subsets of $T$, such that each tuple belongs to exactly
one subset. Each subset of tuples is called a {\bf bucket}. Specifically, let there be $b$ buckets
$B_1,B_2,\ldots,B_b$, then $\cup_{i=1}^bB_i=T$ and for any $1\leq i_1\ne i_2\leq b$, $B_{i_1}\cap
B_{i_2}=\emptyset$.
\end{definition}


\begin{definition}[Slicing]
Given a microdata table $T$, a {\bf slicing} of $T$ is given by an {\bf attribute
partition} and a {\bf tuple partition}.
\end{definition}

For example, Table~\ref{table:example}(e) and Table~\ref{table:example}(f) are two sliced
tables. In Table~\ref{table:example}(e), the attribute partition is \{\{Age\}, \{Sex\},
\{Zipcode\}, \{Disease\}\} and the tuple partition is \{\{$t_1,t_2,t_3,t_4$\},
\{$t_5,t_6, t_7, t_8$\}\}. In Table~\ref{table:example}(f), the attribute partition is
\{\{Age, Sex\}, \{Zipcode, Disease\}\} and the tuple partition is
\{\{$t_1,t_2,t_3,t_4$\}, \{$t_5,t_6,t_7,t_8$\}\}.

Often times, slicing also involves column generalization.

\begin{definition}[Column Generalization]
Given a microdata table $T$ and a column $C_i=\{A_{i1},A_{i2},\ldots,A_{ij}\}$, a {\bf column
generalization} for $C_i$ is defined as a set of non-overlapping $j$-dimensional regions that
\emph{completely} cover $D[A_{i1}]\times D[A_{i2}]\times\ldots\times D[A_{ij}]$. A column
generalization maps each value of $C_i$ to the region in which the value is contained.
\end{definition}

Column generalization ensures that one column satisfies the $k$-anonymity requirement. It is a
multidimensional encoding~\cite{LDR06} and can be used as an additional step in slicing.
Specifically, a general slicing algorithm consists of the following three phases: attribute
partition, column generalization, and tuple partition. Because each column contains much fewer
attributes than the whole table, attribute partition enables slicing to handle high-dimensional
data.

%
%

A key notion of slicing is that of {\em matching buckets}.

\begin{definition}[Matching Buckets]
Let $\{C_1,C_2,\ldots,C_c\}$ be the $c$ columns of a sliced table. Let $t$ be a tuple, and $t[C_i]$
be the $C_i$ value of $t$. Let $B$ be a bucket in the sliced table, and $B[C_i]$ be the multiset of
$C_i$ values in $B$. We say that $B$ is a \emph{matching bucket} of $t$ iff for all $1\leq i\leq
c$, $t[C_i]\in B[C_i]$.
\end{definition}

For example, consider the sliced table shown in Table~\ref{table:example}(f), and consider
$t_1=(22,M,47906,\mathit{dyspepsia})$. Then, the set of matching buckets for $t_1$ is $\{B_1\}$.

\subsection{Comparison with Generalization}
\label{subsection:generalization}

There are several types of recodings for generalization. The recoding that preserves the most
information is \emph{local recoding}.  In local recoding, one first groups tuples into buckets and
then for each bucket, one replaces all values of one attribute with a generalized value. Such a
recoding is local because the same attribute value may be generalized differently when they
appear in different buckets. 

We now show that slicing preserves more information than such a local recoding approach,
assuming that the same tuple partition is used.  We achieve this by showing that
slicing is better than the following enhancement of the local recoding approach.
Rather than using a generalized value to replace more specific attribute values, one uses
%
%
%
the multiset of exact values in each bucket. For example, Table~\ref{table:example}(b) is a
generalized table, and Table~\ref{table:example}(d) is the result of using multisets of exact
values rather than generalized values.
%
%
For the {\em Age} attribute of the first bucket, we use the multiset of exact values
\{22,22,33,52\} rather than the generalized interval $[22-52]$. The multiset of exact values
provides more information about the distribution of values in each attribute than the generalized
interval. Therefore, using multisets of exact values preserves more information than
generalization.

%

However, we observe that this multiset-based generalization is equivalent to a trivial slicing
scheme where each column contains exactly one attribute, because both approaches preserve the exact
values in each attribute but break the association between them within one bucket. For example,
Table~\ref{table:example}(e) is equivalent to Table~\ref{table:example}(d).
Now comparing Table~\ref{table:example}(e) with the sliced table shown in
Table~\ref{table:example}(f), we observe that while one-attribute-per-column slicing preserves
attribute distributional information, it does not preserve attribute correlation, because each
attribute is in its own column.
In slicing, one groups correlated attributes together in one column and preserves their
correlation. For example, in the sliced table shown in Table~\ref{table:example}(f), correlations
between {\em Age} and {\em Sex} and correlations between {\em Zipcode} and {\em Disease} are
preserved. In fact, the sliced table encodes the same amount of information as the original data
with regard to correlations between attributes in the same column.

Another important advantage of slicing is its ability to handle high-dimensional data. By
partitioning attributes into columns, slicing reduces the dimensionality of the data. Each column
of the table can be viewed as a sub-table with a lower dimensionality.
%
%
Slicing is also different from the approach of publishing multiple independent sub-tables in that
these sub-tables are linked by the buckets in slicing.

\subsection{Comparison with Bucketization}
\label{subsection:bucketization}

To compare slicing with bucketization, we first note that bucketization can be viewed as a special
case of slicing, where there are exactly two columns: one column contains only the SA, and the
other contains all the QIs. The advantages of slicing over bucketization can be understood as
follows.
First, by partitioning attributes into more than two columns, slicing can be used to prevent
membership disclosure. Our empirical evaluation on a real dataset shows that bucketization does not
prevent membership disclosure in Section~\ref{sec:experiments}.

Second, unlike bucketization, which requires a clear separation of QI attributes and the sensitive
attribute, slicing can be used without such a separation. For dataset such as the census data, one
often cannot clearly separate QIs from SAs because there is no single external public database
that one can use to determine which attributes the adversary already knows.  Slicing can be useful
for such data.

Finally, by allowing a column to contain both some QI attributes and the sensitive attribute,
attribute correlations between the sensitive attribute and the QI attributes are preserved.  For
example, in Table~\ref{table:example}(f), {\em Zipcode} and {\em Disease} form one column, enabling
inferences about their correlations. Attribute correlations are important utility in data
publishing. For workloads that consider attributes in isolation, one can simply publish two tables,
one containing all QI attributes and one containing the sensitive attribute.

\subsection{Privacy Threats}

When publishing microdata, there are three types of privacy disclosure threats. The first type is
\textit{membership disclosure}. When the dataset to be published is selected from a large
population and the selection criteria are sensitive (e.g., only diabetes patients are selected),
one needs to prevent adversaries from learning whether one's record is included in the published
dataset.

The second type is \textit{identity disclosure}, which occurs when an individual is linked to a
particular record in the released table.  In some situations, one wants to protect against
identity disclosure when the adversary is uncertain of membership.  In this case, protection
against membership disclosure helps protect against identity disclosure.  In other situations,
some adversary may already know that an individual's record is in the published dataset, in which
case, membership disclosure protection either does not apply or is insufficient.

The third type is \textit{attribute disclosure}, which occurs when new information about some
individuals is revealed, i.e., the released data makes it possible to infer the attributes of an
individual more accurately than it would be possible before the release. Similar to the case of
identity disclosure, we need to consider adversaries who already know the membership information.
Identity disclosure leads to attribute disclosure. Once there is identity disclosure, an
individual is re-identified and the corresponding sensitive value is revealed. Attribute
disclosure can occur with or without identity disclosure, e.g., when the sensitive values of all matching tuples are the same.

For slicing, we consider protection against membership disclosure and attribute disclosure. It is a little unclear how identity disclosure should be defined for sliced data (or for data anonymized
by bucketization), since each tuple resides within a bucket and within the bucket the association
across different columns are hidden.  In any case, because identity disclosure leads to attribute
disclosure, protection against attribute disclosure is also sufficient protection against identity
disclosure.

We would like to point out a nice property of slicing that is important for privacy protection. In slicing, a tuple can potentially match multiple buckets, i.e., each tuple can have more than one matching buckets. This is different from previous work on generalization and bucketzation, where each tuple can belong to a unique equivalence-class (or bucket). In fact, it has been recognized~\cite{BS08} that restricting a tuple in a unique bucket helps the adversary but does not improve data utility. We will see that allowing a tuple to match multiple buckets is important for both attribute disclosure protection and attribute disclosure protection, when we describe them in Section~\ref{sec:attribute} and Section~\ref{sec:membership}, respectively.




\section{Attribute Disclosure Protection}
\label{sec:attribute}

In this section, we show how slicing can be used to prevent attribute disclosure, based on the privacy requirement of $\ell$-diversity and introduce the notion of $\ell$-diverse slicing.


\subsection{Example}

We first give an example illustrating how slicing satisfies $\ell$-diversity~\cite{MGK+06} where
the sensitive attribute is ``Disease''.
The sliced table shown in Table~\ref{table:example}(f) satisfies $2$-diversity. Consider tuple
$t_1$ with QI values $(22, M, 47906)$.  In order to determine $t_1$'s sensitive value, one has to
examine $t_1$'s matching buckets.
By examining the first column $(Age,Sex)$ in Table~\ref{table:example}(f), we know that $t_1$ must
be in the first bucket $B_1$ because there are no matches of $(22,M)$ in bucket $B_2$. Therefore, one can conclude that $t_1$ cannot be in bucket $B_2$ and $t_1$ must be in bucket $B_1$.

Then, by examining the $Zipcode$ attribute of the second column $(Zipcode, Disease)$ in bucket
$B_1$, we know that the column value for $t_1$ must be either $(47906,dyspepsia)$ or $(47906,flu)$
because they are the only values that match $t_1$'s zipcode 47906. Note that the other two column
values have zipcode 47905.
Without additional knowledge, both {\em dyspepsia} and {\em flu} are equally possible to be the
sensitive value of $t_1$. Therefore, the probability of learning the correct sensitive value of
$t_1$ is bounded by $0.5$.
Similarly, we can verify that $2$-diversity is satisfied for all other tuples in Table~\ref{table:example}(f).

\subsection{$\ell$-Diverse Slicing}
\label{subsec:diversity}

In the above example, tuple $t_1$ has only one matching bucket. In general, a tuple $t$ can have
multiple matching buckets. We now extend the above analysis to the general case and introduce the
notion of $\ell$-diverse slicing.

Consider an adversary who knows all the QI values of $t$ and attempts to infer $t$'s sensitive
value from the sliced table. She or he first needs to determine which buckets $t$ may reside in,
i.e., the set of matching buckets of $t$.
Tuple $t$ can be in any one of its matching buckets. Let $p(t,B)$ be the
probability that $t$ is in bucket $B$ (the procedure for computing $p(t,B)$ will be described later
in this section). For example, in the above example, $p(t_1,B_1)=1$ and $p(t_1,B_2)=0$.

In the second step, the adversary computes $p(t,s)$, the probability that $t$ takes a sensitive value $s$.
$p(t,s)$ is calculated using {\em the law of total probability}. Specifically, let $p(s|t,B)$ be
the probability that $t$ takes sensitive value $s$ given that $t$ is in bucket $B$, then according to the law of total probability, the
probability $p(t,s)$ is:

\begin{equation}
p(t,s)=\sum_{B}p(t,B)p(s|t,B)
\label{eq:law}
\end{equation}

In the rest of this section, we show how to compute the two probabilities: $p(t,B)$ and $p(s|t,B)$.

\mypara{Computing $p(t,B)$.} Given a tuple $t$ and a sliced bucket $B$, the probability that $t$ is
in $B$ depends on the fraction of $t$'s column values that match the column values in $B$.
If some column value of $t$ does not appear in the corresponding column of $B$, it is certain that
$t$ is not in $B$.
In general, bucket $B$ can potentially match $|B|^c$ tuples, where $|B|$ is the number of tuples in
$B$. Without additional knowledge, one has to assume that the column values are independent;
therefore each of the $|B|^c$ tuples is equally likely to be an original tuple. The probability
that $t$ is in $B$ depends on the fraction of the $|B|^c$ tuples that match $t$.

We formalize the above analysis. We consider the match between $t$'s column values $\{t[C_1],
t[C_2], \cdots, t[C_c]\}$ and $B$'s column values $\{B[C_1], B[C_2], \cdots, B[C_c]\}$.
Let $f_i(t,B)$ $(1\leq i\leq c-1)$ be the fraction of occurrences of $t[C_i]$ in $B[C_i]$ and let
$f_c(t, B)$ be the fraction of occurrences of $t[C_c-\{S\}]$ in $B[C_c-\{S\}])$.
Note that, $C_c-\{S\}$ is the set of QI attributes in the sensitive column. For example, in
Table~\ref{table:example}(f), $f_1(t_1,B_1)=1/4=0.25$ and $f_2(t_1,B_1)=2/4=0.5$. Similarly,
$f_1(t_1,B_2)=0$ and $f_2(t_1,B_2)=0$.
Intuitively, $f_i(t,B)$ measures the {\em matching degree} on column $C_i$, between tuple $t$ and bucket $B$.

Because each possible candidate tuple is equally likely to be an original tuple, the {\em
matching degree} between $t$ and $B$ is the product of the matching degree on
each column, i.e., $f(t,B)=\prod_{1\leq i\leq c}f_i(t,B)$.
Note that $\sum_{t}f(t,B)=1$ and when $B$ is not a matching bucket of $t$, $f(t,B)=0$.

Tuple $t$ may have multiple matching buckets, $t$'s total matching degree in the whole data is
$f(t)=\sum_{B}f(t,B)$. The probability that $t$ is in bucket $B$ is:
\[p(t,B)=\frac{f(t,B)}{f(t)}\]

\mypara{Computing $p(s|t,B)$.} Suppose that $t$ is in bucket $B$, to determine $t$'s sensitive
value, one needs to examine the sensitive column of bucket $B$. Since the sensitive column contains
the QI attributes, not all sensitive values can be $t$'s sensitive value. Only those sensitive
values whose QI values match $t$'s QI values are $t$'s {\em candidate sensitive values}. Without
additional knowledge, all candidate sensitive values (including duplicates) in a bucket are equally
possible. Let $D(t,B)$ be the distribution of $t$'s candidate sensitive values in bucket $B$.

\begin{definition}[$D(t,B)$]
Any sensitive value that is associated with $t[C_c-\{S\}]$ in $B$ is a {\bf candidate sensitive
value} for $t$ (there are $f_c(t,B)$ candidate sensitive values for $t$ in $B$, including
duplicates). Let $D(t,B)$ be the distribution of the candidate sensitive values in $B$ and
$D(t,B)[s]$ be the probability of the sensitive value $s$ in the distribution.
\end{definition}

For example, in Table~\ref{table:example}(f), $D(t_1,B_1)=(dyspepsia: 0.5, flu: 0.5)$ and therefore
$D(t_1,B_1)[dyspepsia]=0.5$.
The probability $p(s|t,B)$ is exactly $D(t,B)[s]$, i.e., $p(s|t,B)=D(t,B)[s]$.

\mypara{$\ell$-Diverse Slicing.} Once we have computed $p(t,B)$ and $p(s|t,B)$, we are able to
compute the probability $p(t,s)$ based on the Equation (\ref{eq:law}).
We can show when $t$ is in the data, the probabilities that $t$ takes a sensitive value sum up to
$1$.
\begin{fact}
For any tuple $t\in D$, $\sum_{s}p(t,s)=1$.
\end{fact}
\begin{proof}
\begin{equation}
\begin{split}
\sum_{s}p(t,s)& = \sum_{s}\sum_{B}p(t,B)p(s|t,B)\\
& = \sum_{B}p(t,B)\sum_{s}p(s|t,B)\\
& = \sum_{B}p(t,B)\\
& = 1\\
\end{split}
\end{equation}
\end{proof}

$\ell$-Diverse slicing is defined based on the probability $p(t,s)$.
\begin{definition}[$\ell$-diverse slicing]
A tuple $t$ satisfies $\ell$-diversity iff for any sensitive value $s$,
\[p(t,s)\leq 1/\ell\]
A sliced table satisfies $\ell$-diversity iff every tuple in it satisfies $\ell$-diversity.
\end{definition}

Our analysis above directly show that from an $\ell$-diverse sliced table, an adversary cannot
correctly learn the sensitive value of any individual with a probability greater than $1/\ell$.
Note that once we have computed the probability that a tuple takes a sensitive value, we can also use slicing for other privacy measures such as $t$-closeness~\cite{LLV07}.

%

\section{Slicing Algorithms}
\label{sec:algorithms}

We now present an efficient slicing algorithm to achieve $\ell$-diverse slicing.
Given a microdata table $T$ and two parameters $c$ and $\ell$, the algorithm computes the sliced
table that consists of $c$ columns and satisfies the privacy requirement of $\ell$-diversity.

Our algorithm consists of three phases: {\em attribute partitioning}, {\em column generalization},
and {\em tuple partitioning}. 
We now describe the three phases.

\subsection{Attribute Partitioning}

Our algorithm partitions attributes so that highly-correlated attributes are in the same column.
This is good for both utility and privacy. In terms of data utility, grouping highly-correlated
attributes preserves the correlations among those attributes. In terms of privacy, the association
of uncorrelated attributes presents higher identification risks than the association of
highly-correlated attributes because the association of uncorrelated attribute values is much less
frequent and thus more identifiable. Therefore, it is better to break the associations between
uncorrelated attributes, in order to protect privacy.

In this phase, we first compute the correlations between pairs of attributes and then cluster
attributes based on their correlations.

\subsubsection{Measures of Correlation}
\label{subsubsec:measures}

%

Two widely-used measures of association are Pearson correlation coefficient~\cite{Cra48} and
mean-square contingency coefficient~\cite{Cra48}. Pearson correlation coefficient is used for
measuring correlations between two continuous attributes while mean-square contingency coefficient
is a chi-square measure of correlation between two categorical attributes.
We choose to use the {\em mean-square contingency coefficient} because most of our attributes are
categorical. Given two attributes $A_1$ and $A_2$ with domains $\{v_{11},v_{12},...,v_{1d_1}\}$ and
$\{v_{21},v_{22},...,v_{2d_2}\}$, respectively. Their domain sizes are thus $d_1$ and $d_2$,
respectively.
The mean-square contingency coefficient between $A_1$ and $A_2$ is defined as:

\[\phi^2(A_1,A_2)=\frac{1}{\min\{d_1,d_2\}-1}\sum_{i=1}^{d_1}\sum_{j=1}^{d_2}\frac{(f_{ij}-f_{i\cdot}f_{\cdot
j})^2}{f_{i \cdot}f_{\cdot j}}\]

Here, $f_{i\cdot}$ and $f_{\cdot j}$ are the fraction of occurrences of $v_{1i}$ and $v_{2j}$ in
the data, respectively. $f_{ij}$ is the fraction of co-occurrences of $v_{1i}$ and $v_{2j}$ in the
data. Therefore,  $f_{i\cdot}$ and $f_{\cdot j}$ are the marginal totals of $f_{ij}$:
$f_{i\cdot}=\sum_{j=1}^{d_2}f_{ij}$ and $f_{\cdot j}=\sum_{i=1}^{d_1}f_{ij}$.
It can be shown that $0\leq \phi^2(A_1,A_2)\leq 1$.

%
For continuous attributes, we first apply {\em discretization} to partition the domain of a
continuous attribute into intervals and then treat the collection of interval values as a discrete
domain. Discretization has been frequently used for decision tree classification, summarization,
and frequent itemset mining.
%
%
We use equal-width discretization, which partitions an attribute domain into (some $k$) equal-sized
intervals. Other methods for handling continuous attributes are the subjects of future work.
%
%

\subsubsection{Attribute Clustering}
\label{subsubsec:clustering}

Having computed the correlations for each pair of attributes, we use clustering to partition
attributes into columns. In our algorithm, each attribute is a point in the clustering space.  The
distance between two attributes in the clustering space is defined as
$d(A_1,A_2)=1-\phi^2(A_1,A_2)$, which is in between of $0$ and $1$. Two attributes that are
strongly-correlated will have a smaller distance between the corresponding data points in our
clustering space.

We choose the $k$-medoid method for the following reasons. First, many existing clustering
algorithms (e.g., $k$-means) requires the calculation of the ``centroids''. But there is no notion
of ``centroids'' in our setting where each attribute forms a data point in the clustering space.
Second, $k$-medoid method is very robust to the existence of outliers (i.e., data points that are
very far away from the rest of data points). Third, the order in which the data points are examined
does not affect the clusters computed from the $k$-medoid method. We use the well-known $k$-medoid
algorithm PAM (Partition Around Medoids)~\cite{KR90}. PAM starts by an arbitrary selection of $k$
data points as the initial medoids. In each subsequent step, PAM chooses one medoid point and one
non-medoid point and swaps them as long as the cost of clustering decreases. Here, the clustering
cost is measured as the sum of the cost of each cluster, which is in turn measured as the sum of
the distance from each data point in the cluster to the medoid point of the cluster.
The time complexity of PAM is $O(k(n-k)^2)$. Thus, it is known that PAM suffers from high
computational complexity for large datasets. However, the data points in our clustering space are
attributes, rather than tuples in the microdata. Therefore, PAM will not have computational
problems for clustering attributes.

\subsubsection{Special Attribute Partitioning}

In the above procedure, all attributes (including both QIs and SAs) are clustered into columns. The
$k$-medoid method ensures that the attributes are clustered into $k$ columns but does not have any
guarantee on the size of the sensitive column $C_c$. In some cases, we may pre-determine the number
of attributes in the sensitive column to be $\alpha$.
The parameter $\alpha$ determines the size of the sensitive column $C_c$, i.e., $|C_c|=\alpha$.
If $\alpha=1$, then $|C_c|=1$, which means that $C_c=\{S\}$. And when $c=2$, slicing in this case
becomes equivalent to bucketization.
If $\alpha>1$, then $|C_c|>1$, the sensitive column also contains some QI attributes.

We adapt the above algorithm to partition attributes into $c$ columns such that the sensitive
column $C_c$ contains $\alpha$ attributes.
We first calculate correlations between the sensitive attribute $S$ and each QI attribute.  Then, we rank the QI
attributes by the decreasing order of their correlations with $S$ and select the top $\alpha-1$ QI
attributes. Now, the sensitive column $C_c$ consists of $S$ and the selected QI attributes. All
other QI attributes form the other $c-1$ columns using the attribute clustering algorithm.

\subsection{Column Generalization}

In the second phase, tuples are generalized to satisfy some minimal frequency requirement. We want
to point out that column generalization is not an indispensable phase in our algorithm. As shown by
Xiao and Tao~\cite{XT06b}, bucketization provides the same level of privacy protection as
generalization, with respect to attribute disclosure.

Although column generalization is not a required phase, it can be useful in several aspects. First,
column generalization may be required for identity/membership disclosure protection. If a column
value is unique in a column (i.e., the column value appears only once in the column), a tuple with
this unique column value can only have one matching bucket. This is not good for privacy
protection, as in the case of generalization/bucketization where each tuple can belong to only
one equivalence-class/bucket. The main problem is that this unique column value can be identifying.
In this case, it would be useful to apply column generalization to ensure that each column value
appears with at least some frequency.

\begin{figure}[t]
\centering
\begin{tabular}{l}
{\bf Algorithm tuple-partition($T,\ell$)}\\
1. $Q=\{T\}$; $\SB=\emptyset$.\\
2. while $Q$ is not empty\\
3. \hspace*{0.5cm}remove the first bucket $B$ from $Q$; $Q=Q-\{B\}$.\\
4. \hspace*{0.5cm}split $B$ into two buckets $B_1$ and $B_2$, as in Mondrian.\\
5. \hspace*{0.5cm}if {\bf diversity-check}($T$, $Q\cup\{B_1,B_2\}\cup\SB$, $\ell$)\\
6. \hspace*{1.0cm}$Q=Q\cup\{B_1,B_2\}$.\\
7. \hspace*{0.5cm}else $\SB=\SB\cup\{B\}$.\\
8. return $\SB$.\\
\end{tabular}
\caption{The tuple-partition algorithm}
\vspace*{-.2in}
\label{fig:algorithm}
\end{figure}

Second, when column generalization is applied, to achieve the same level of privacy against attribute disclosure, bucket sizes can be smaller (see Section~\ref{subsec:tuple}). While column generalization may result in information loss, smaller bucket-sizes allows better data utility. Therefore, there is a trade-off between column generalization and tuple partitioning.
In this paper, we mainly focus on the tuple partitioning algorithm. The tradeoff between column generalization and tuple partitioning is the subject of future work.
Existing anonymization algorithms can be used for column generalization, e.g., Mondrian~\cite{LDR06}. The algorithms can be
applied on the sub-table containing only attributes in one column to ensure the anonymity
requirement.

\subsection{Tuple Partitioning}
\label{subsec:tuple}

In the tuple partitioning phase, tuples are partitioned into buckets. We modify the
Mondrian~\cite{LDR06} algorithm for tuple partition. Unlike Mondrian $k$-anonymity, no
generalization is applied to the tuples; we use Mondrian for the purpose of partitioning tuples
into buckets.

Figure~\ref{fig:algorithm} gives the description of the tuple-partition algorithm. The algorithm
maintains two data structures: (1) a queue of buckets $Q$ and (2) a set of sliced buckets $\SB$.
Initially, $Q$ contains only one bucket which includes all tuples and $\SB$ is empty (line 1).
In each iteration (line 2 to line 7), the algorithm removes a bucket from $Q$ and splits the bucket
into two buckets (the split criteria is described in Mondrian~\cite{LDR06}). If the sliced table
after the split satisfies $\ell$-diversity (line 5), then the algorithm puts the two buckets at the
end of the queue $Q$ (for more splits, line 6). Otherwise, we cannot split the bucket anymore and
the algorithm puts the bucket into $\SB$ (line 7). When $Q$ becomes empty, we have computed the
sliced table. The set of sliced buckets is $\SB$ (line 8).

The main part of the tuple-partition algorithm is to check whether a sliced table satisfies
$\ell$-diversity (line 5). Figure~\ref{fig:diversity} gives a description of the {\em diversity-check} algorithm.
For each tuple $t$, the algorithm maintains a list of statistics $L[t]$ about $t$'s matching
buckets. Each element in the list $L[t]$ contains statistics about one matching bucket $B$: the matching probability $p(t,B)$ and the distribution of candidate sensitive values $D(t,B)$.

\begin{figure}[t]
\centering
\begin{tabular}{l}
{\bf Algorithm diversity-check($T,T^*,\ell$)}\\
1. for each tuple $t\in T$, $L[t]=\emptyset$.\\
2. for each bucket $B$ in $T^*$\\
3. \hspace*{0.5cm}record $f(v)$ for each column value $v$ in bucket $B$.\\
4. \hspace*{0.5cm}for each tuple $t\in T$\\
5. \hspace*{1.0cm}calculate $p(t,B)$ and find $D(t,B)$.\\
6. \hspace*{1.0cm}$L[t]=L[t]\cup\{\tuple{p(t,B),D(t,B)}\}$.\\
7. for each tuple $t\in T$\\
8. \hspace*{0.5cm}calculate $p(t,s)$ for each $s$ based on $L[t]$.\\
9. \hspace*{0.5cm}if $p(t,s)\geq 1/\ell$, return false.\\
10. return true.
\end{tabular}
\caption{The diversity-check algorithm}
\vspace*{-.2in}
\label{fig:diversity}
\end{figure}

The algorithm first takes one scan of each bucket $B$ (line 2 to line 3) to record the frequency $f(v)$ of each column value $v$ in bucket $B$. Then the algorithm takes one scan of each tuple $t$ in the table $T$ (line 4 to line 6) to
find out all tuples that match $B$ and record their matching probability $p(t,B)$ and the
distribution of candidate sensitive values $D(t,B)$, which are added to the list $L[t]$ (line 6).
At the end of line 6, we have obtained, for each tuple $t$, the list of statistics $L[t]$
about its matching buckets. A final scan of the tuples in $T$ will compute the $p(t,s)$ values
based on {\em the law of total probability} described in Section~\ref{subsec:diversity}. Specifically,
\[p(t,s)=\sum_{e\in L[t]}e.p(t,B)* e.D(t,B)[s]\]
The sliced table is $\ell$-diverse iff for all sensitive value $s$, $p(t,s)\leq 1/\ell$ (line 7 to line 10).

We now analyze the time complexity of the tuple-partition algorithm. The time complexity of
Mondrian~\cite{LDR06} or kd-tree~\cite{FBF77} is $O(n\log n)$ because at each level of the kd-tree,
the whole dataset need to be scanned which takes $O(n)$ time and the height of the tree is $O(\log
n)$. In our modification, each level takes $O(n^2)$ time because of the diversity-check algorithm
(note that the number of buckets is at most $n$). The total time complexity is therefore $O(n^2\log
n)$.

\section{Membership Disclosure Protection}
\label{sec:membership}

Let us first examine how an adversary can infer membership information from bucketization. Because
bucketization releases the QI values in their original form and most individuals can be uniquely
identified using the QI values, the adversary can simply determine the membership of an individual
in the original data by examining the frequency of the QI values in the bucketized data.
Specifically, if the frequency is 0, the adversary knows for sure that the individual is not in the data.
If the frequency is greater than 0, the adversary knows with high confidence that the individual is
in the data, because this matching tuple must belong to that individual as almost no other
individual has the same QI values.

The above reasoning suggests that in order to protect membership information, it is required that, in the anonymized data, a tuple in the original data should have a similar frequency as a tuple that is not in the original data.
Otherwise, by examining their frequencies in the anonymized data, the adversary can differentiate
tuples in the original data from tuples not in the original data.

We now show how slicing protects against membership disclosure.
Let $D$ be the set of tuples in the original data and let $\overline{D}$ be the set of tuples that are
not in the original data. Let $D^s$ be the sliced data. Given
$D^s$ and a tuple $t$, the goal of membership disclosure is to determine whether $t\in D$ or
$t\in \overline{D}$.
In order to distinguish tuples in $D$ from tuples in $\overline{D}$, we examine their differences. If $t\in D$, $t$ must have at least one matching buckets in $D^s$. To protect membership
information, we must ensure that at least some tuples in $\overline{D}$ should also have matching
buckets. Otherwise, the adversary can differentiate between $t\in D$ and $t\in \overline{D}$ by
examining the number of matching buckets.

We call a tuple {\em an original tuple} if it is in $D$. We call a tuple
{\em a fake tuple} if it is in $\overline{D}$ and it matches at least one bucket in the
sliced data.
Therefore, we have considered two measures for membership disclosure protection. The first measure is the number of fake tuples. When the number of fake tuples is 0 (as in bucketization), the membership information of every tuple can be determined.
The second measure is to consider the number of matching buckets for original tuples and that for fake tuples. If they are similar enough, membership information is protected because the adversary cannot distinguish original tuples from fake tuples.

Slicing is an effective technique for membership disclosure protection. A sliced bucket of size $k$
can potentially match $k^c$ tuples. Besides the original $k$ tuples, this bucket can introduce as
many as $k^c-k$ tuples in $\overline{D}$, which is $k^{c-1}-1$ times more than the number of
original tuples.
The existence of such tuples in $\overline{D}$ hides the membership information of tuples in $D$,
because when the adversary finds a matching bucket, she or he is not certain whether this tuple is
in $D$ or not since a large number of tuples in $\overline{D}$ have matching buckets as well.
In our experiments
(Section~\ref{sec:experiments}), we empirically evaluate slicing in membership disclosure protection.

\section{Experiments}
\label{sec:experiments}

We conduct two experiments.
In the first experiment, we evaluate the effectiveness of slicing in preserving data utility and protecting against attribute disclosure, as
compared to generalization and bucketization.
To allow direct comparison, we use the Mondrian algorithm~\cite{LDR06} and $\ell$-diversity for all three anonymization techniques: generalization, bucketization, and slicing.
This experiment demonstrates that: (1) slicing preserves better data utility than generalization;
(2) slicing is more effective than bucketization in workloads involving the sensitive attribute;
and (3) the sliced table can be computed efficiently.
Results for this experiment are presented in Section~\ref{subsection:exp-attribute}.

In the second experiment, we show the effectiveness of slicing in membership disclosure
protection. For this purpose, we count the number of fake tuples in the sliced data.
We also compare the number of matching buckets for original tuples and that for fake tuples.
Our experiment results show that bucketization does not prevent membership disclosure as almost
every tuple is uniquely identifiable in the bucketized data. Slicing provides better protection
against membership disclosure: (1) the number of fake tuples in the sliced data is very large, as compared to the number of original tuples and (2) the number of matching buckets for fake tuples and that for original tuples are close enough, which makes it difficult for the adversary to distinguish fake tuples from original tuples. Results for this experiment are presented in Section~\ref{subsec:exp-membership}.

\begin{table}[t]
\centering
\begin{tabular}{|c|c|c|c|}
\hline &Attribute&Type&\# of values\\
\hline 1&Age&Continuous&74\\
\hline 2&Workclass&Categorical&8\\
\hline 3&Final-Weight&Continuous&NA\\
\hline 4&Education&Categorical&16\\
\hline 5&Education-Num&Continuous&16\\
\hline 6&Marital-Status&Categorical&7\\
\hline 7&Occupation&Categorical&14\\
\hline 8&Relationship&Categorical&6\\
\hline 9&Race&Categorical&5\\
\hline 10&Sex&Categorical&2\\
\hline 11&Capital-Gain&Continuous&NA\\
\hline 12&Capital-Loss&Continuous&NA\\
\hline 13&Hours-Per-Week&Continuous&NA\\
\hline 14&Country&Categorical&41\\
\hline 15&Salary&Categorical&2\\
\hline
\end{tabular}
\caption{Description of the \emph{\textbf{Adult}} dataset} \label{tab:description} 
\vspace*{-.2in}
\end{table}

\mypara{Experimental Data.} We use the Adult dataset from the UC Irvine machine learning
repository~\cite{AN07}, which is comprised of data collected from the US census. The dataset is described in
Table~\ref{tab:description}. Tuples with missing values are eliminated and there are $45222$ valid
tuples in total.
The adult dataset contains $15$ attributes in total.

In our experiments, we obtain two datasets from the Adult dataset. The first dataset is the
``OCC-7'' dataset, which includes $7$ attributes: $QI=$ $\{Age,$ $Workclass,$ $Education,$
$Marital$-$Status,$ $Race,$ $Sex\}$ and $S=Occupation$. The second dataset is the ``OCC-15''
dataset, which includes all
$15$ attributes and the sensitive attribute is $S=Occupation$. 
%

In the ``OCC-7'' dataset, the attribute that has the closest correlation with the sensitive
attribute {\em Occupation} is {\em Gender}, with the next closest attribute being {\em Education}.
In the ``OCC-15'' dataset, the closest attribute is also {\em Gender} but the next closest
attribute is {\em Salary}.

\subsection{Preprocessing}
Some preprocessing steps must be applied on the anonymized data before it can be used for workload
tasks. First, the anonymized table computed through generalization contains generalized values,
which need to be transformed to some form that can be understood by the classification algorithm.
Second, the anonymized table computed by bucketization or slicing contains multiple columns, the
linking between which is broken. We need to process such data before workload experiments can run
on the data.

\mypara{Handling generalized values.} In this step, we map the generalized values (set/interval) to
data points.
Note that the Mondrian algorithm assumes a total order on the domain values of each attribute and
each generalized value is a sub-sequence of the total-ordered domain values. There are several
approaches to handle generalized values. The first approach is to replace a generalized value with
the {\em mean} value of the generalized set. For example, the generalized age [20,54] will be
replaced by age 37 and the generalized Education level \{9th,10th,11th\} will be replaced by 10th.
The second approach is to replace a generalized value by its lower bound and upper bound. In this
approach, each attribute is replaced by two attributes, doubling the total number of attributes.
For example, the Education attribute is replaced by two attributes {\em Lower-Education} and {\em
Upper-Education}; for the generalized Education level \{9th, 10th, 11th\}, the {\em
Lower-Education} value would be 9th and the {\em Upper-Education value} would be 11th.
For simplicity, we use the second approach in our experiments.

\mypara{Handling bucketized/sliced data.} In both bucketization and slicing, attributes are
partitioned into two or more columns. For a bucket that contains $k$ tuples and $c$ columns, we
generate $k$ tuples as follows. We first randomly permutate the values in each column. Then, we
generate the $i$-th ($1\leq i\leq k$) tuple by linking the $i$-th value in each column. We apply
this procedure to all buckets and generate all of the tuples from the bucketized/sliced table. This
procedure generates the linking between the two columns in a random fashion. In all of our
classification experiments, we applies this procedure $5$ times and the average results are
reported.

\begin{figure}
\centering
\begin{tabular}{cc}
\begin{minipage}{1.6in}
\centering
{\includegraphics[height=1.4in,width=1.5in]{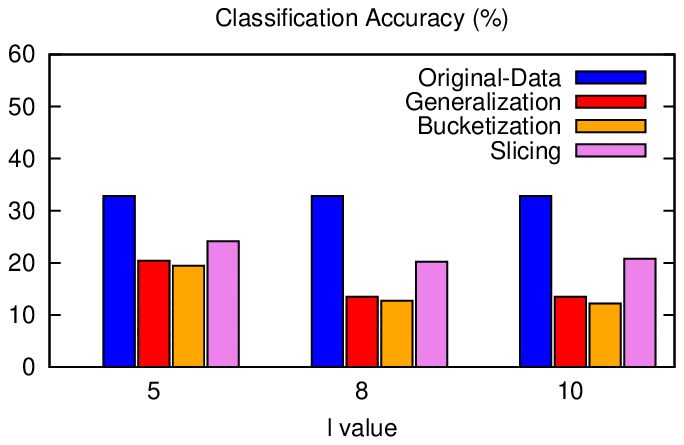}}
\end{minipage}
&
\begin{minipage}{1.6in}
\centering
{\includegraphics[height=1.4in,width=1.5in]{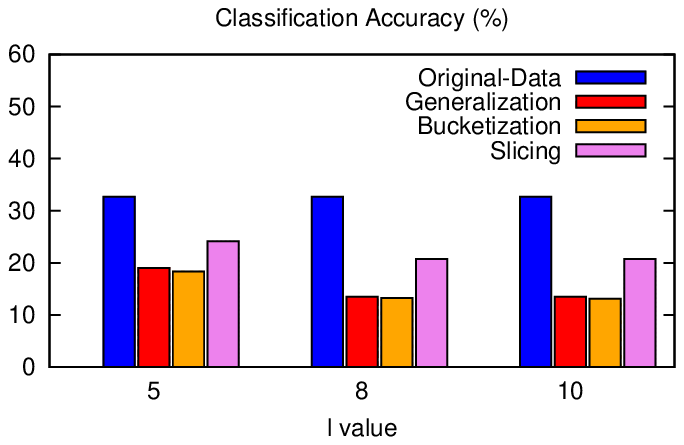}}
\end{minipage}
\vspace*{.1in}
\\
\hspace{0.2cm}(a) J48 (OCC-7) & (b) Naive Bayes (OCC-7)
\\
\begin{minipage}{1.6in}
\centering
{\includegraphics[height=1.4in,width=1.5in]{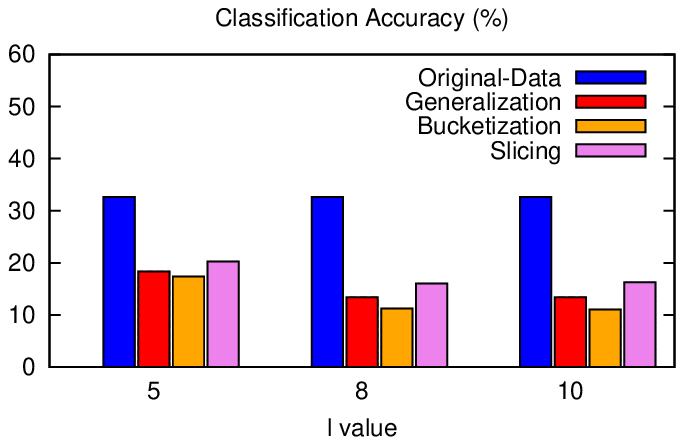}}
\end{minipage}
&
\begin{minipage}{1.6in}
\centering
{\includegraphics[height=1.4in,width=1.5in]{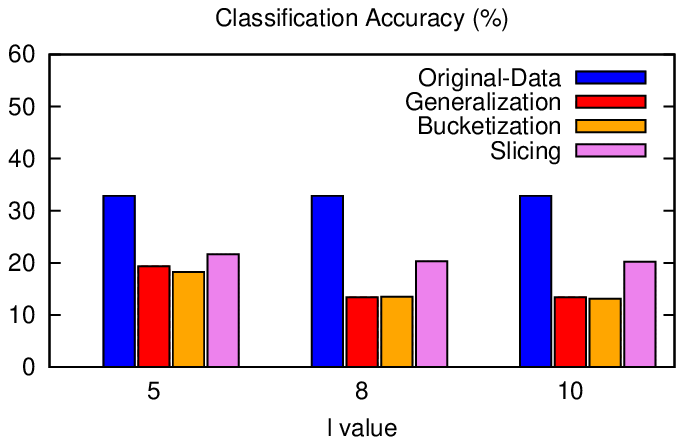}}
\end{minipage}
\vspace*{.1in}
\\
\hspace{0.2cm}(c) J48 (OCC-15) & (d) Naive Bayes (OCC-15)
\end{tabular}
\caption{Learning the sensitive attribute (Target: Occupation)}
\vspace*{-.2in}
\label{fig:sensitive}
\end{figure}

\subsection{Attribute Disclosure Protection}
\label{subsection:exp-attribute}

We compare slicing with generalization and bucketization on data utility of the anonymized data for
classifier learning. For all three techniques, we employ the Mondrian algorithm~\cite{LDR06} to
compute the $\ell$-diverse tables. The $\ell$ value can take values \{5,8,10\} (note that the {\em
Occupation} attribute has 14 distinct values).
In this experiment, we choose $\alpha=2$. Therefore, the sensitive column is always
{\em\{Gender, Occupation\}}.

\mypara{Classifier learning.} We evaluate the quality of the anonymized data for classifier
learning, which has been used in~\cite{FWY05,LDR06b,BS08}. We use the Weka software package to
evaluate the classification accuracy for Decision Tree C4.5 (J48) and Naive Bayes. Default settings are used in both tasks. For all classification experiments, we use 10-fold cross-validation.

In our experiments, we choose one attribute as the target attribute (the attribute on which the
classifier is built) and all other attributes serve as the predictor attributes.
We consider the performances of the anonymization algorithms in both
learning the sensitive attribute {\em Occupation} and learning a QI attribute {\em Education}.

\mypara{Learning the sensitive attribute.} In this experiment, we build a classifier on the
sensitive attribute, which is ``{\em Occupation}''. We fix $c=2$ here and evaluate the effects of $c$ later in this section. 
%
Figure~\ref{fig:sensitive} compares the quality of the anonymized data (generated by the three
techniques) with the quality of the original data, when the target attribute is {\em Occupation}.
The experiments are performed on the two datasets OCC-7 (with 7 attributes) and OCC-15 (with 15
attributes).
%

In all experiments, slicing outperforms both generalization and bucketization, that confirms that
slicing preserves attribute correlations between the sensitive attribute and some QIs (recall that
the sensitive column is {\em\{Gender, Occupation\}}). Another observation is that bucketization
performs even slightly worse than generalization. That is mostly due to our preprocessing step that
randomly associates the sensitive values to the QI values in each bucket. This may introduce false
associations while in generalization, the associations are always correct although the exact
associations are hidden. A final observation is that when $\ell$ increases, the performances of
generalization and bucketization deteriorate much faster than slicing. This also confirms that
slicing preserves better data utility in workloads involving the sensitive attribute.

\begin{figure}[t]
\centering
\begin{tabular}{cc}
\begin{minipage}{1.6in}
\centering
{\includegraphics[height=1.4in,width=1.5in]{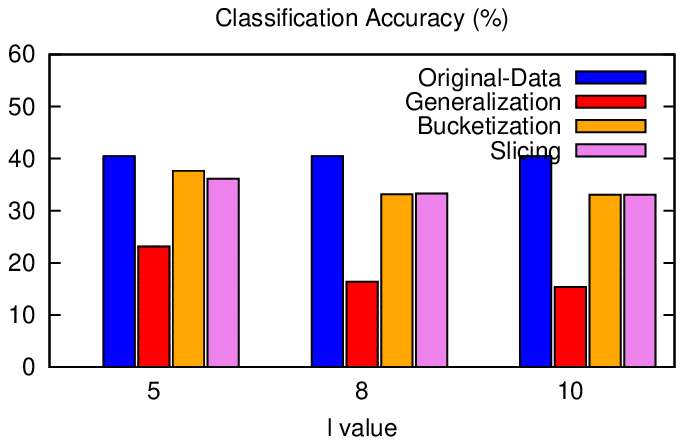}}
\end{minipage}
&
\begin{minipage}{1.6in}
\centering
{\includegraphics[height=1.4in,width=1.5in]{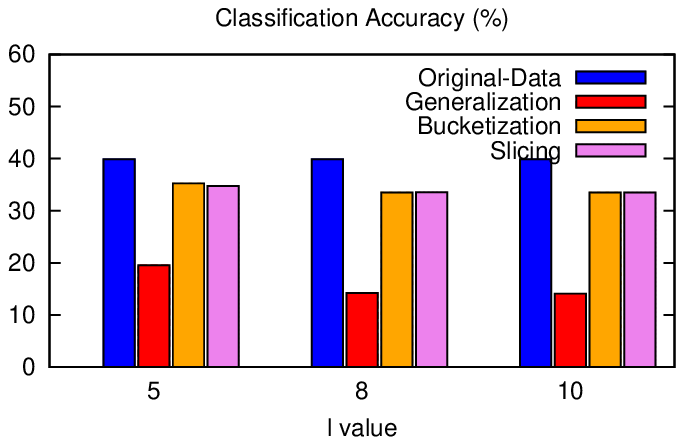}}
\end{minipage}
\vspace*{.1in}
\\
\hspace{0.2cm}(a) J48 (OCC-7) & (b) Naive Bayes (OCC-7)
\\
\begin{minipage}{1.6in}
\centering
{\includegraphics[height=1.4in,width=1.5in]{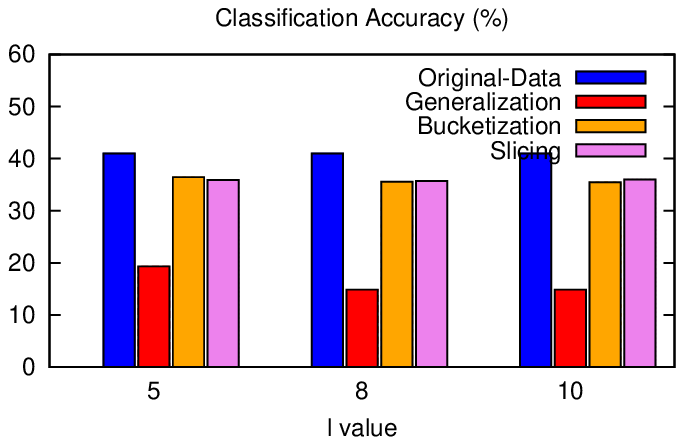}}
\end{minipage}
&
\begin{minipage}{1.6in}
\centering
{\includegraphics[height=1.4in,width=1.5in]{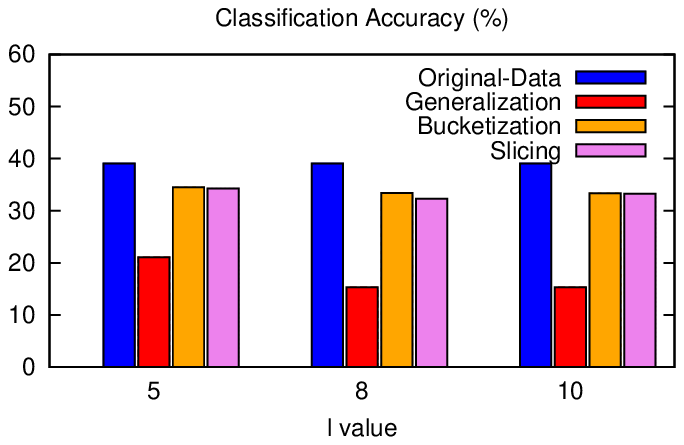}}
\end{minipage}
\vspace*{.1in}
\\
\hspace{0.2cm}(c) J48 (OCC-15) & (d) Naive Bayes (OCC-15)
\end{tabular}
\caption{Learning a QI attribute (Target: Education)}
\label{fig:QI}
\vspace*{-.2in}
\end{figure}

\mypara{Learning a QI attribute.} In this experiment, we build a classifier on the QI
attribute ``{\em Education}''. We fix $c=2$ here and evaluate the effects of $c$ later in this section. 
%
Figure~\ref{fig:QI} shows the experiment results.

In all experiments, both bucketization and slicing perform much better than generalization. This is
because in both bucketization and slicing, the QI attribute {\em Education} is in the same column
with many other QI attributes: in bucketization, all QI attributes are in the same column; in
slicing, all QI attributes except {\em Gender} are in the same column. This fact allows both
approaches to perform well in workloads involving the QI attributes.
Note that the classification accuracies of bucketization and slicing are lower than that of the
original data. This is because the sensitive attribute {\em Occupation} is closely correlated with
the target attribute {\em Education} (as mentioned earlier in Section~\ref{sec:experiments}, {\em
Education} is the second closest attribute with {\em Occupation} in OCC-7). By breaking the link
between {\em Education} and {\em Occupation}, classification accuracy on {\em Education} reduces
for both bucketization and slicing.

\begin{figure}[t]
\centering
\begin{tabular}{cc}
\begin{minipage}{1.6in}
\centering
{\includegraphics[height=1.4in,width=1.5in]{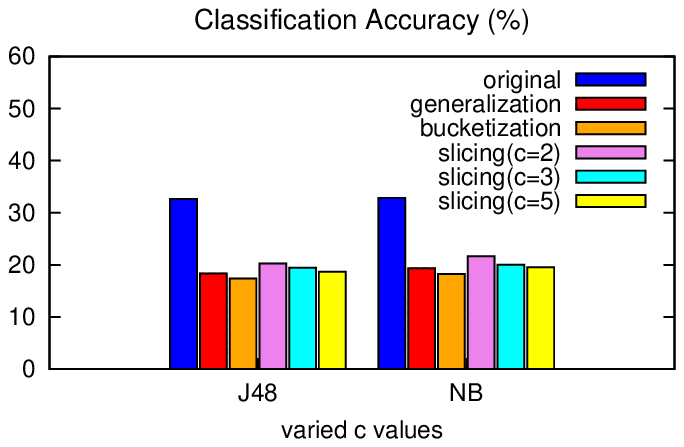}}
\end{minipage}
&
\begin{minipage}{1.6in}
\centering
{\includegraphics[height=1.4in,width=1.5in]{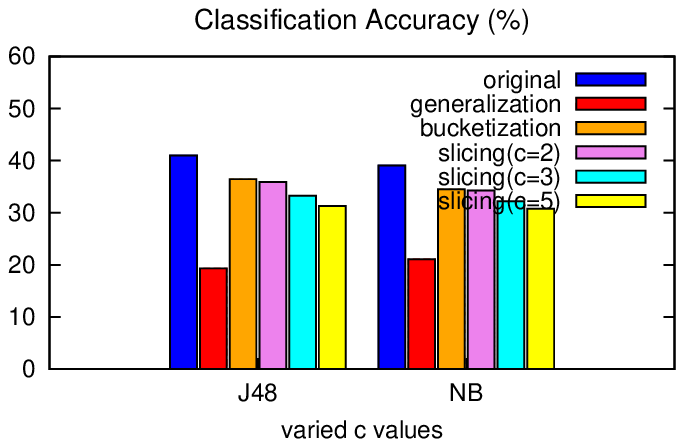}}
\end{minipage}
\vspace*{.1in}
\\
\hspace{0.2cm}(a) Sensitive (OCC-15) & (b) QI (OCC-15)
\end{tabular}
\caption{Varied $c$ values}
\label{fig:C}
\vspace*{-.2in}
\end{figure}

\mypara{The effects of $c$.} In this experiment, we evaluate the effect of $c$ on classification accuracy. We fix $\ell=5$ and vary the number of columns $c$ in \{2,3,5\}. Figure~\ref{fig:C}(a) shows the results on learning the sensitive attribute and Figure~\ref{fig:C}(b) shows the results on learning a QI attribute. It can be seen that classification accuracy decreases only slightly when we increase $c$, because the most correlated attributes are still in the same column.
In all cases, slicing shows better accuracy than generalization. When the target attribute is the sensitive attribute, slicing even performs better than bucketization.

\subsection{Membership Disclosure Protection}
\label{subsec:exp-membership}

In the second experiment, we evaluate the effectiveness of slicing in membership disclosure protection.

We first show that bucketization is vulnerable to membership disclosure. In both the OCC-7 dataset
and the OCC-15 dataset, each combination of QI values occurs exactly once. This means that the adversary can determine the membership information of any individual by checking if the QI value appears in the bucketized data. If the QI value does not appear in the bucketized data, the individual is not in the original data. Otherwise, with high confidence, the individual is in the original data as no other individual has the same QI value.

We then show that slicing does prevent membership disclosure. We perform  the following experiment. %
First, we partition attributes into $c$ columns based on attribute correlations. We set $c\in
\{2,5\}$. In other words, we compare $2$-column-slicing with $5$-column-slicing.
For example, when we set $c=5$, we obtain $5$ columns. In OCC-7, \{$\mathit{Age}$,
$\mathit{Marriage}$, $\mathit{Gender}$\} is one column and each other attribute is in its own
column. In OCC-15, the $5$ columns are: \{$\mathit{Age}$, $\mathit{Workclass}$,
$\mathit{Education}$, $\mathit{Education}$-$\mathit{Num}$, $\mathit{Cap}$-$\mathit{Gain}$,
$\mathit{Hours}$, $\mathit{Salary}$\}, \{$\mathit{Marriage}$, $\mathit{Occupation}$,
$\mathit{Family}$, $\mathit{Gender}$\}, \{$\mathit{Race}$,$\mathit{Country}$\},
\{$\mathit{Final}$-$\mathit{Weight}$\}, and \{$\mathit{Cap}$-$\mathit{Loss}$\}.

Then, we randomly partition tuples into buckets of size $p$ (the last bucket may have fewer than
$p$ tuples). As described in Section~\ref{sec:membership}, we collect statistics about the following two measures in our experiments: (1) the number of fake tuples and (2) the number of matching buckets for original v.s. the number of matching buckets for fake tuples.

\begin{figure}[t]
\centering
\begin{tabular}{cc}
\begin{minipage}{1.6in}
\centering
{\includegraphics[height=1.4in,width=1.5in]{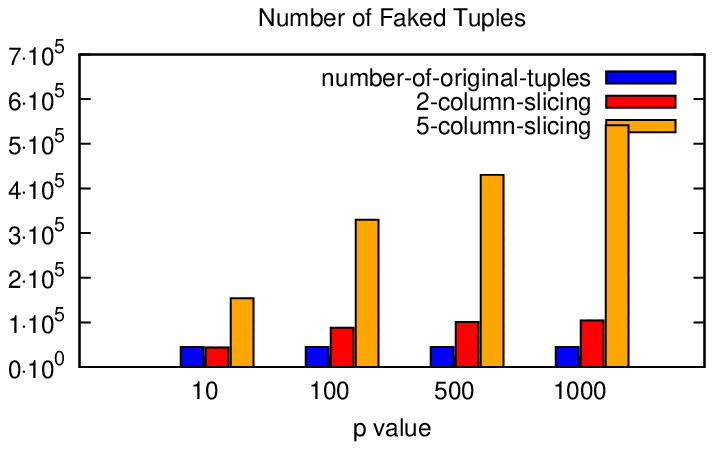}}
\end{minipage}
&
\begin{minipage}{1.6in}
\centering
{\includegraphics[height=1.4in,width=1.5in]{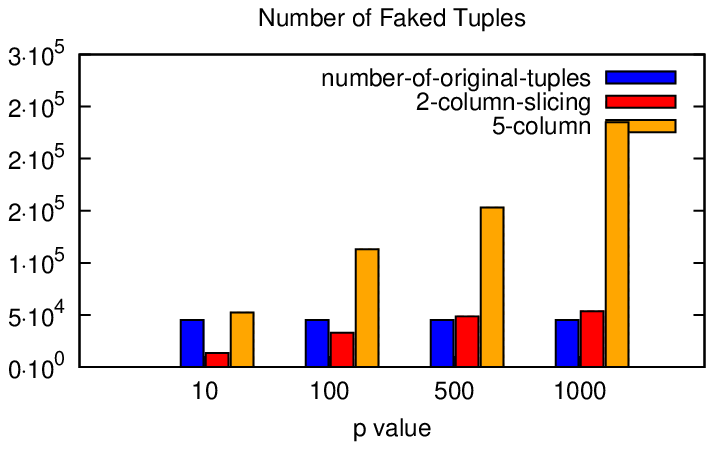}}
\end{minipage}
\vspace*{.1in}
\\
\hspace{0.2cm}(a) OCC-7 & (b) OCC-15
\end{tabular}
\caption{Number of fake tuples}
\label{fig:number}
\vspace*{-.2in}
\end{figure}

\mypara{The number of fake tuples.}
Figure~\ref{fig:number} shows the experimental results on the number of fake tuples, with respect to
the bucket size $p$. 
%
Our results show that the number of fake tuples is large enough to hide the original tuples.
For example, for the OCC-7 dataset, even for a small bucket size of $100$ and only $2$ columns, slicing introduces as many as $87936$ fake tuples, which is
nearly twice the number of original tuples ($45222$).
When we increase the bucket size, the number of fake tuples becomes larger. This is consistent
with our analysis that a bucket of size $k$ can potentially match $k^c-k$ fake tuples.
In particular, when we increase the number of columns $c$, the number of fake tuples becomes
exponentially larger. In almost all experiments, the number of fake tuples is larger than the
number of original tuples.
The existence of such a large number of fake tuples provides protection for membership information
of the original tuples.

\begin{figure}[t]
\centering
\begin{tabular}{cc}
\begin{minipage}{1.6in}
\centering
{\includegraphics[height=1.4in,width=1.5in]{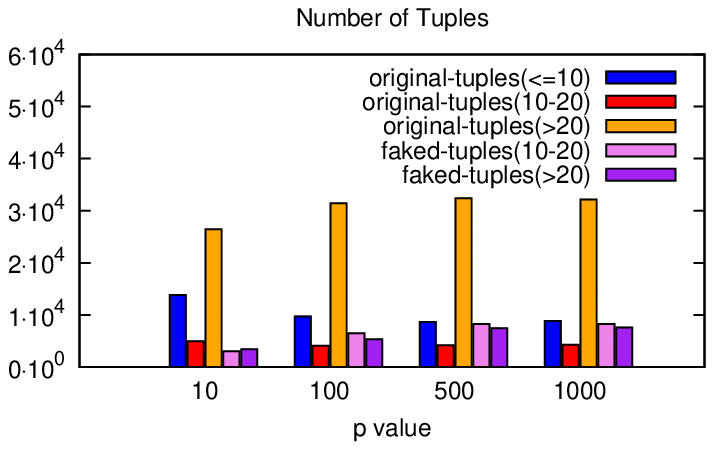}}
\end{minipage}
&
\begin{minipage}{1.6in}
\centering
{\includegraphics[height=1.4in,width=1.5in]{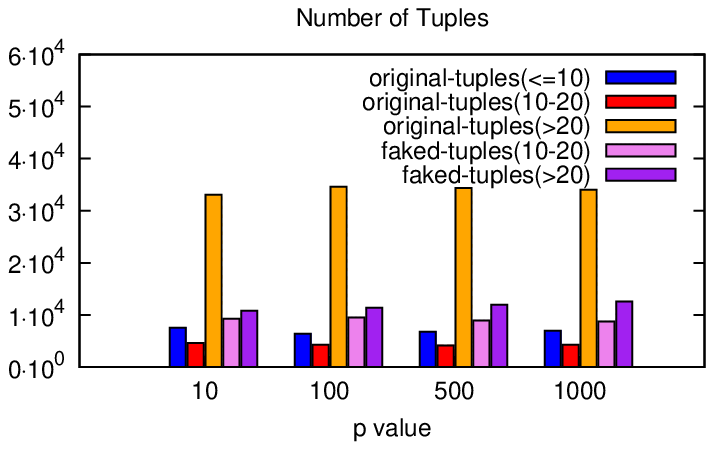}}
\end{minipage}
\vspace*{.1in}
\\
\hspace{0.2cm}(a) $2$-column (OCC-7) & (b) $5$-column (OCC-7)
\\
\begin{minipage}{1.6in}
\centering
{\includegraphics[height=1.4in,width=1.5in]{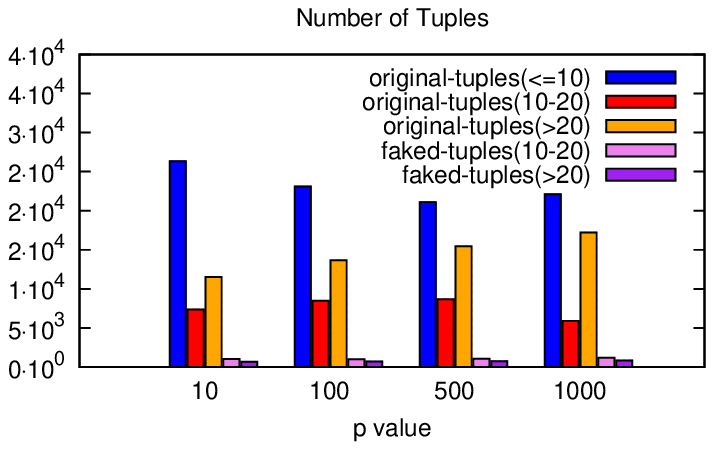}}
\end{minipage}
&
\begin{minipage}{1.6in}
\centering
{\includegraphics[height=1.4in,width=1.5in]{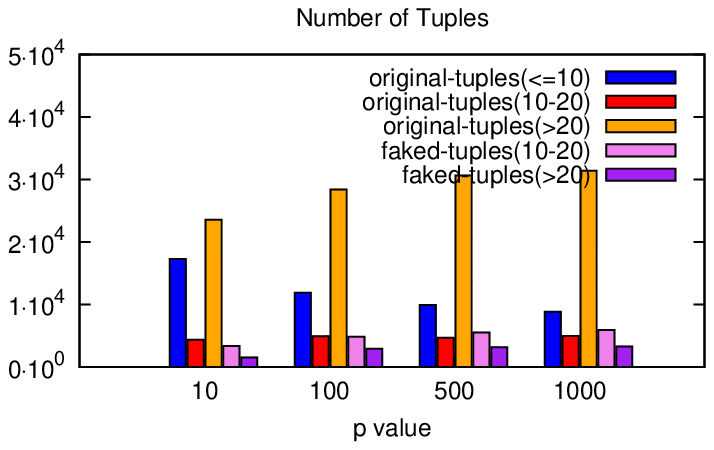}}
\end{minipage}
\vspace*{.1in}
\\
\hspace{0.2cm}(c) $2$-column (OCC-15) & (d) $5$-column (OCC-15)
\end{tabular}
\caption{Number of tuples that have matching buckets}
\label{fig:matching}
\vspace*{-.2in}
\end{figure}

\mypara{The number of matching buckets.} Figure~\ref{fig:matching} shows the number of matching buckets for original tuples and fake
tuples. 
%

We categorize the tuples (both original tuples and fake tuples) into three categories: (1) $\leq 10$:
tuples that have at most 10 matching buckets, (2) $10-20$: tuples that have more than 10 matching
buckets but at most 20 matching buckets, and (3) $>20$: tuples that have more than 20 matching
buckets.
For example, the ``original-tuples($\leq 10$)'' bar gives the number of original tuples that have
at most 10 matching buckets and the ``fake-tuples($>20$)'' bar gives the number of fake tuples
that have more than $20$ matching buckets. Because the number of fake tuples that have at most 10
matching buckets is very large, we omit the ``fake-tuples($\leq 10$)'' bar from the figures to
make the figures more readable.

Our results show that, even when we do random grouping, many fake tuples have a large number of
matching buckets. For example, for the OCC-7 dataset, for a small $p=100$ and $c=2$, there are
$5325$ fake tuples that have more than $20$ matching buckets; the number is $31452$ for original tuples.
The numbers are even closer for larger $p$ and $c$ values.
This means that a larger bucket size and more columns provide better
protection against membership disclosure.

Although many fake tuples have a large number of matching buckets, in general, original tuples
have more matching buckets than fake tuples. As we can see from the figures, a large fraction of
original tuples have more than $20$ matching buckets while only a small fraction of fake tuples
have more than $20$ tuples. This is mainly due to the fact that we use random grouping in the
experiments.
The results of random grouping are that the number of fake tuples is very large but most fake
tuples have very few matching buckets. When we aim at protecting membership information, we can
design more effective grouping algorithms to ensure better protection against membership
disclosure. The design of tuple grouping algorithms
is left to future work.

\section{Related Work}
\label{sec:related}

Two popular anonymization techniques are generalization and bucketization.
Generalization~\cite{Sam01,Swe02a,Swe02b} replaces a value with a ``less-specific but
semantically consistent'' value. Three types of  encoding schemes have been proposed for
generalization: global recoding, regional recoding, and local recoding.
Global recoding has the property that multiple occurrences of the same value are always replaced by
the same generalized value.
Regional record~\cite{LDR06} is also called multi-dimensional recoding (the Mondrian algorithm)
which partitions the domain space into non-intersect regions and data points in the same region are
represented by the region they are in.
Local recoding does not have the above constraints and allows different occurrences of the same
value to be generalized differently.

Bucketization~\cite{XT06b,MKM+07,KSY+07} first partitions tuples in the table into buckets and then
separates the quasi-identifiers with the sensitive attribute by randomly permuting the sensitive
attribute values in each bucket. The anonymized data consists of a set of buckets with permuted
sensitive attribute values.
%
%
In particular, bucketization has been used for anonymizing high-dimensional data~\cite{GTK08}.
Please refer to Section~\ref{subsection:generalization} and Section~\ref{subsection:bucketization}
for a detailed comparison of slicing with generalization and bucketization, respectively.

Slicing has some connections to marginal publication~\cite{KG06}; both of them
release correlations among a subset of attributes.
Slicing is quite different from marginal publication in a number of aspects. 
First, marginal publication can be viewed as a special case of slicing which does not 
have horizontal partitioning. Therefore, correlations among attributes in different columns 
are lost in marginal publication. By horizontal partitioning, attribute correlations between 
different columns (at the bucket level) are preserved. Marginal publication is similar to 
overlapping vertical partitioning, which is left as our future work (See Section~\ref{sec:conclusions}).
Second, the key idea of 
slicing is to preserve correlations between highly-correlated attributes 
and to break correlations between uncorrelated attributes, thus achieving 
both better utility and better privacy. Third, existing data analysis 
(e.g., query answering) methods can be easily used on the sliced data.

Existing privacy measures for membership disclosure protection include differential
privacy~\cite{DN03,Dwo06,DMN+06} and $\delta$-presence~\cite{NAC07}. Differential privacy has
recently received much attention in data privacy, especially for interactive databases~\cite{DN03,BDM+05,Dwo06,DMN+06,XT08}.  Rastogi et al.~\cite{RSH07} design the
$\alpha\beta$ algorithm for data perturbation that satisfies differential privacy. Machanavajjhala
et al.~\cite{MKA+08} apply the notion of differential privacy for synthetic data generation.
On the other hand, $\delta$-presence~\cite{NAC07} assumes that the published database is a sample
of a large public database and the adversary has knowledge of this large database. The calculation
of disclosure risk depends on this large database.

Finally, privacy measures for attribute disclosure protection include
$\ell$-diversity~\cite{MGK+06}, $(\alpha,k)$-anonymity~\cite{WLF+06}, $t$-closeness~\cite{LLV07},
$(k,e)$-anonymity~\cite{KSY+07}, $(c,k)$-safety~\cite{MKM+07}, privacy skyline~\cite{CRL07},
$m$-confidentiality~\cite{WFW+07} and $(\epsilon,m)$-anonymity~\cite{LTX08}. We use
$\ell$-diversity in slicing for attribute disclosure protection.

\section{Discussions and Future Work}
\label{sec:conclusions}
This paper presents a new approach called slicing to privacy-preserving microdata publishing.
Slicing overcomes the limitations of generalization and bucketization and preserves better utility
while protecting against privacy threats.
We illustrate how to use slicing to prevent attribute disclosure and membership disclosure. Our
experiments show that slicing preserves better data utility than generalization and is more
effective than bucketization in workloads involving the sensitive attribute.

The general methodology proposed by this work is that: before anonymizing the data, one can analyze
the data characteristics and use these characteristics in data anonymization. The rationale is that
one can design better data anonymization techniques when we know the data better. In~\cite{LL08}, we show that 
attribute correlations can be used for privacy attacks.

This work motivates several directions for future research. First, in this paper, we consider slicing where each attribute is in exactly one column. An extension is the
notion of {\em overlapping slicing}, which duplicates an attribute in more than one columns. This
releases more attribute correlations. For example, in Table~\ref{table:example}(f),
one could choose to include the {\em Disease} attribute also in the first column. That is, the two
columns are $\{Age, Sex, Disease\}$ and $\{Zipcode, Disease\}$. This could provide better data
utility, but the privacy implications need to be carefully studied and understood. It is interesting to study
the tradeoff between privacy and utility~\cite{LL09}.

Second, we plan to study membership disclosure protection in more details. Our experiments show
that random grouping is not very effective. We plan to design more effective tuple grouping algorithms. 

Third, slicing is a promising technique for handling high-dimensional data. By partitioning attributes into columns, we protect privacy by breaking the
association of uncorrelated attributes and preserve data utility by preserving the association
between highly-correlated attributes. For example, slicing can be used for anonymizing transaction databases, which has been studied recently in~\cite{TMK08,XWF+08,NS08}.

Finally, while a number of anonymization techniques have been designed, it remains an open problem
on how to use the anonymized data. In our experiments, we randomly generate the associations
between column values of a bucket. This may lose data utility. Another direction to design data
mining tasks using the anonymized data~\cite{IKB09} computed by various anonymization techniques.

%
%

\bibliographystyle{abbrv}
\bibliography{Privacy}

\end{document}